\documentclass[twocolumn,a4paper,fleqn]{article}
\usepackage{epsfig}% SRC Specials

\newcommand{\be}{\begin{equation}}
\newcommand{\ee}{\end{equation}}
\newcommand{\ba}{\begin{eqnarray}}
\newcommand{\ea}{\end{eqnarray}}
\newcommand{\ban}{\begin{eqnarray*}}
\newcommand{\ean}{\end{eqnarray*}}
\begin{document}

\title{\Large\bf Relativistic Nonlocality \\
in experiments with successive impacts}

{\normalsize{\author{{\bf Antoine Suarez}\thanks{suarez@leman.ch}\\
Center for Quantum Philosophy\\ The Institute for Interdisciplinary
Studies\\ P.O. Box 304, CH-8044 Zurich, Switzerland }}}
%\date{}
\maketitle

\begin{abstract}
Relativistic Nonlocality is applied to experiments in which one of
the photons impacts successively at two beam-splitters. It is
discussed whether a time series with 2 {\em non-before} impacts can
be produced with beam-splitters at rest and such an experiment may
allow us to decide between Quantum Mechanics (QM) and Relativistic
Nonlocality (RNL).\\

{\em Keywords:}  relativistic nonlocality, multisimultaneity,
timing-depen\-dent joint probabilities, 2 {\em non-before} impacts.

\end{abstract}

\section{Introduction}

Relativistic Nonlocality (RNL) is an alternative nonlocal
description which unifies the relativity of simultaneity and
superluminal nonlocality, avoiding superluminal signaling. Its main
feature is Multisimultaneity, i.e. each particle at the time it
inpacts on a beam-splitter, in te referential frame of this beam
splitter, takes account of what happens to the other "entangled"
particles. Multisimultaneity implies rules to calculate joint
probabilities which are unknown in QM, and deviates from the time
insensitivity of the QM formalism: In RNL which rule applies to
calculate probabilities depends not only on indistinguishability
but also on the timing of the impacts at the beam-splitters
\cite{as97.2, asvs97.2}.\\

In previous articles RNL has been applied to experiments with fast
moving beam-splitters. As well for experiments with 2 {\em before}
impacts \cite{asvs97.1}, as for such with 2 {\em non-before}
impacts \cite{as97.2} RNL leads to predictions conflicting with
QM.\\

The possibility of testing time insensitivity with beam-splitters
at rest has also been suggested \cite{as97.1}. In this article we
explore more in depth this possibility. In an experiment in which
one of the particles impacts successively at two beam-splitters
before getting detected, three different time series can be
arranged, one of them exhibiting 2 {\em non-before} impacts. It is
argued that for this case results contradicting QM cannot be
excluded, and therefore it may be a profitable endeavour to perform
the corresponding experiment.

\section{Experiments with photons impacting successively
at two beam-splitters}

%%%%%%%%%%%%%%%%%%%%%%%%%%%%%%
\begin{figure}[b]
\begin{center}
\epsfig{figure=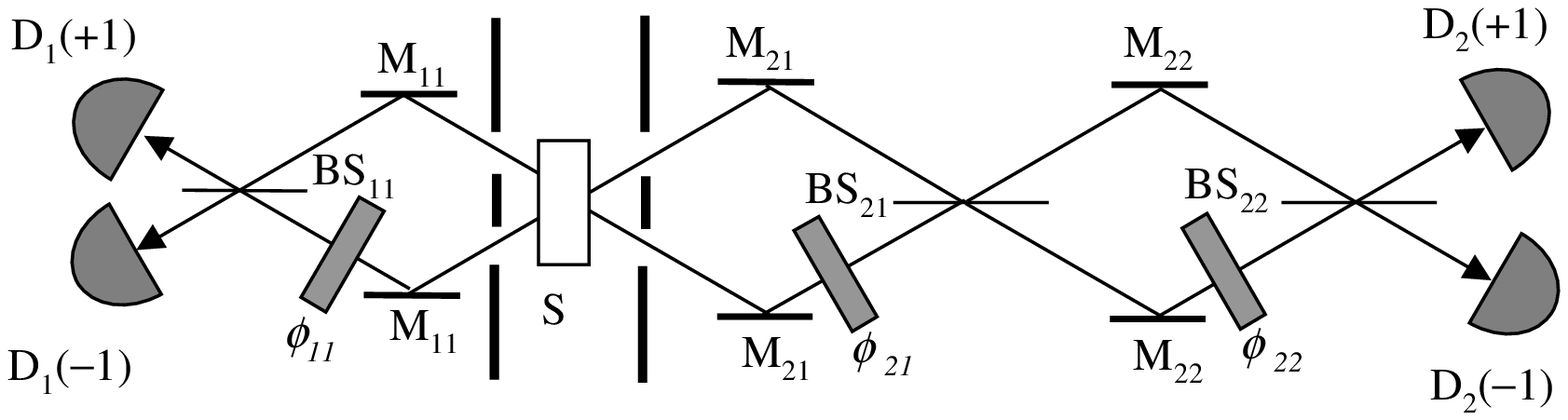,width=76mm}
{\small\it{\caption{Experiment with photon 2 impacting successively
at resting beam-splitters BS$_{21}$ and BS$_{22}$.}}}
\label{fig:setup1}
\end{center}
\end{figure}
%%%%%%%%%%%%%%%%%%%%%%%%%%%%%%

Consider the gedankenexperiment represented in Fig. 1. Two photons
emitted back-to-back in a "Bell state", can travel by alternative
pairs of paths from the source S to either one of the left-hand
detectors D$_{1}(+1)$ and D$_{1}(-1)$ and either one of the
rigt-hand detectors D$_{2}(+1)$ and D$_{2}(-1)$. Before they are
getting detected photon 1 impacts on beam-splitter BS$_{11}$, and
photon 2 impacts successively on beam-splitters BS$_{21}$ and
BS$_{22}$. The phase parameters are labeled $\phi_{11}$,
$\phi_{21}$ and $\phi_{22}$. The beam-splitters are supposed at
rest in the laboratory frame.\\

By displacing the mirrors M$_{11}$ it is possible to achieve three
different Time Series in the laboratory frame:

\begin{enumerate}
\item{The impact on BS$_{22}$ occurs before the impact BS$_{11}$.}
\item{The impact on BS$_{11}$ occurs before the impact on BS$_{21}$.}
\item{The impact on BS$_{21}$ occurs before the impact on
BS$_{11}$the impact on BS$_{11}$ occurs before the impact on
BS$_{22}$.}
\end{enumerate}

Unless stated otherwise, we assume in the following these two {\em
indistinguishability} conditions:\\

{\em Condition 1}: Through detection of photon 1 after BS$_{11}$
and detection of photon 2 between BS$_{21}$ and B$_{22}$ it is in
principle impossible to know to which input sub-ensemble a particle
pair belongs.\\

{\em Condition 2}: Through detection of photon 1 after BS$_{11}$
and detection of photon 2 after BS$_{22}$ it is in principle
impossible to know which path photon 2 did travel, neither before
its arrival at BS$_{21}$, nor before its arrival at BS$_{22}$.\\

In the following sections we discuss the three Time Series
considered above, first according to QM and thereafter according to
RNL

\section{The QM description}

The conventional application of the quantum mechanical
superposition principle considers all three time series as being
equivalent. The relative time ordering of the impacts at the
beam-splitters does not influence the distribution of the outcomes;
in this respect only indistinguishability matters: if it is
impossible to obtain path information the
sum-of-probability-amplitudes rule applies. Accordingly for all
three time series QM predicts:\\

\ba
P^{QM}(u_{11},u_{22})_{\sigma\omega}\nonumber\\
=\frac{1}{4}+\frac{\sigma\omega}{8}\Big(cos(\phi_{11}-\phi_{21}-\phi_{22})\nonumber\\
-cos(\phi_{11}-\phi_{21}+\phi_{22})\Big),
\label{eq:Pu11u22}
\ea

where $\sigma,\omega\in\{+,-\}$, and
$P^{QM}(u_{11},u_{22})_{\sigma\omega}$ denote the quantum
mechanical joint probabilities for the four possible outcomes
obtained through detections after BS$_{11}$ and BS$_{22}$ under the
indistinguishability {\em condition 2}. From Eq. (\ref{eq:Pu11u22})
follows the correlation coefficient:

\ba
E^{QM}=\sum_{\sigma,\omega}\sigma\omega
P^{QM}(u_{11},u_{22})_{\sigma\omega}\nonumber\\
=\frac{1}{2}\Big(cos(\phi_{11}-\phi_{21}-\phi_{22})\nonumber\\
-cos(\phi_{11}-\phi_{21}+\phi_{22})\Big).
\label{eq:Eu11u22}
\ea

\smallskip

\section{The RNL description}
\label{se:RNL}

The basic principles and theorems of RNL presented in \cite{as97.2}
are now extended to experiments with successive impacts. We discuss
experiments with moving beam-splitters involving multisimultaneity
(i.e. several simultaneity frames) and, as particular cases, the
three possible time series in the experiment of Fig. 1 with
beam-splitters at rest (i.e., involving only one simultaneity
frame).\\

At time $T_{ik}$ at which particle $i$, $(i\in\{1,2\})$, arrives at
beam-splitter BS$_{ik}$ we consider in the inertial frame of this
beam-splitter which beam-splitters BS$_{jl}$ particle $j$,
$(j\in\{1,2\}, j\neq i)$ did already reach, i.e. we consider
whether the relation $(T_{ik}<T_{j1})_{ik}$ holds, or there is a
BS$_{jl}$ such that the relation $(T_{jl}\leq
T_{ik}<T_{jl+1})_{ik}$ holds, the subscript $ik$ after the
parenthesis meaning that all times referred to are measured in the
inertial frame of BS$_{ik}$.

\subsection{Timing $(b_{11}$, $b_{22})$}

If $(T_{11}<T_{21})_{11}$, then we consider the impact on BS$_{11}$
to be a {\em before} one, and we label it $b_{11}$.

If $(T_{21}<T_{11})_{21}$, we consider the impact on BS$_{21}$ to
be a {\em before} one, and label it $b_{21}$.

If $(T_{22}<T_{11})_{22}$ and $(T_{21}<T_{11})_{21}$, then we
assume the impact on BS$_{22}$ to be a {\em before} one, and we
label it  $b_{22})$. However, if $(T_{22}<T_{11})_{22}$, but
$(T_{21}\geq T_{11})_{21}$, the impact on BS$_{22}$ would be a {\em
non-before} one.\\

{\em Principle I} of RNL implies:

\ba
P(b_{11},b_{21})_{\sigma\omega}=P^{QM}(d_{11},d_{21})_{\sigma\omega}=\frac{1}{4},
\label{eq:Pb11b21}
\ea

where $P^{QM}(d_{11},d_{21})_{\sigma\omega}$ denotes the joint
probabilities predicted by standard QM if the particles are
detected after BS$_{11}$ and BS$_{21}$, and it is possible to know
which path photon $i$ travels before entering BS$_{i1}$, i.e., to
which of the two prepared sub-ensembles the photon pair belongs.\\

Eq. (\ref{eq:Pb11b21}) leads to the correlation coefficient:

\ba
E(b_{11},b_{21})=\sum_{\sigma,\omega}\sigma\omega
P(b_{11},b_{21})_{\sigma\omega}\nonumber\\
=\frac{1}{4}\sum_{\sigma,\omega}\sigma\omega=0.
\label{eq:Eb11b21}
\ea

Similarly, we assume that the photons of a pair undergoing impacts
$b_{11}$ and $b_{22}$ produce values taking into account only local
information, i.e., photon $i$ does not become influenced by the
parameters photon $j$ meets at the other arm of the setup.
Therefore {\em Principle I} of RNL implies that:

\ba
P(b_{11},b_{22})_{\sigma\omega}=P^{QM}(d_{11},d_{22})_{\sigma\omega}=\frac{1}{4},
\label{eq:Pb11b22}
\ea

where $P^{QM}(d_{11},d_{22})_{\sigma\omega}$ denotes the joint
probabilities predicted by standard QM if the particles are
detected after BS$_{11}$ and BS$_{22}$, and it is possible to know
which polarization photon $i$ has before entering BS$_{i1}$, i.e.,
to which of the two prepared sub-ensemble the photon pair
belongs.\\

Accordingly one is led to the correlation coefficient:

\ba
E(b_{11},b_{22})=\sum_{\sigma,\omega}\sigma\omega
P(b_{11},b_{22})_{\sigma\omega}\nonumber\\
=\frac{1}{4}\sum_{\sigma,\omega}\sigma\omega=0.
\label{eq:Eb11b22}
\ea

\smallskip

\subsection{Timing $(a_{11[22]}$,$b_{22})$ (e.g. Series 1),
and $(b_{11}$, $a_{22})$ (e.g. Series 2)}

If $(T_{22}>T_{11}\geq T_{21})_{11}$, we assume the impact on
BS$_{11}$ to be a {\em non-before} one with relation to the impact
on BS$_{21}$, and label it as $a_{11[21]}$. If $(T_{11}\geq
T_{22})_{11}$, we assume the impact on BS$_{11}$ to be a {\em
non-before} one with relation to the impacts on BS$_{22}$, and
label it as $a_{11[22]}$.\\

Similarly, if $(T_{21}\geq T_{11})_{21}$, or $(T_{22}\geq
T_{11})_{22}$, we assume the impact on BS$_{22}$ to be a {\em
non-before} one with relation to the impacts on BS$_{11}$, and we
label it $a_{22[11]}$, or simply $a_{22}$ since no ambiguity
results.\\

First af all consider an experiment $(a_{11[21]}$, $b_{21})$ in
which the photons are detected after leaving BS$_{11}$ and
BS$_{21}$. As stated in \cite{as97.2} ({\em Principle II}) RNL
considers the correlations to reveal causal links, and assumes the
values $(a_{11[21]})_{\sigma}$ to depend on the values
$(b_{21})_{\omega}$ as follows:

\ba
P(a_{11[21]},b_{21})_{\sigma\omega}
=P^{QM}(u_{11},u_{21})_{\sigma\omega},
\label{eq:Pa11[21]b21}
\ea

what yields the correlation coefficient:

\ba
E(a_{11[21]},b_{21})=\cos (\phi_{11}-\phi_{21}).
\label{eq:Ea11[21]b21}
\ea

Principle (\ref{eq:Pa11[21]b21}) can be extended straighforward to
experiments $(a_{11[22]}$, $b_{22})$ and $(b_{11}$, $a_{22})$ as
follows:

\ba
P(a_{11[22]},b_{22})_{\sigma\omega}
=P(b_{11},a_{22})_{\sigma\omega}\nonumber\\
=P^{QM}(u_{11},u_{22})_{\sigma\omega}.
\label{eq:Pa11[22]b22}
\ea

Obviously, {\em time series 1} corresponds to an experiment
$(a_{11[22]}$, $b_{22})$, and {\em time series 2} to a $(b_{11}$,
$a_{22})$ one, and therefore, taking Eq. (\ref{eq:Eu11u22}) into
account, one is led to the following correlation coefficient:

\ba
E(a_{11[22]},b_{22})=E(b_{11},a_{22})\nonumber\\
=\frac{1}{2}\Big(cos(\phi_{11}-\phi_{21}-\phi_{22})\nonumber\\
-cos(\phi_{11}-\phi_{21}+\phi_{22})\Big).
\label{eq:Ea11[22]b22}
\ea

Eq. (\ref{eq:Ea11[22]b22}) and the preceding Eq. (\ref{eq:Eb11b22})
can be considered the translation into mathematical terms of Bell's
claim: "Correlations cry out for explanation".

\subsection{Timing $(a_{11[22]}$, $a_{22})$: Need for conditional probabilities}

We consider now an experiment in which the impact on BS$_{11}$ is
{\em non-before} with relation to the impact on BS$_{22}$, and the
impact on BS$_{22}$ is {\em non-before} with relation to the impact
on BS$_{11}$. As discussed in \cite{as97.2}, it would be absurd to
assume together that the impacts on BS$_{22}$ take into account the
outcomes of the impacts on BS$_{11}$, and the impacts on BS$_{11}$
take into account the outcomes of the impacts on BS$_{22}$. That is
why RNL assumes that photon $i$ undergoing an $a_{ik[jl]}$ impact
always takes account of the values $(b_{jl})_{\omega}$ photon $j$
had produced in a {\em before} impact, but not necessarily of the
values $(a_{jl[ik']})_{\omega'}$ photon $j$ actually produces.\\

To put this principle into an equation requires the introduction of
conditional probabilities. We denote by
$P\Big((a_{ik[jl]})_{\sigma'}|(b_{ik},b_{jl})_{\sigma\omega}\Big)$
the probability that a particle pair that would have produced the
outcome $({\sigma,\omega})$ in a $(b_{ik},b_{jl})$ experiment,
produces the outcome $({\sigma',\omega})$ if the experiment is a
$(a_{ik[jl]},b_{jl})$ one. Then it holds that:

\ba
P(a_{11[22]},a_{22})_{\sigma'\omega'}
=\sum_{\sigma,\omega}P(b_{11},b_{22})_{\sigma\omega}\nonumber\\
\times P\Big((a_{11[22]})_{\sigma'}|(b_{11},b_{22})_{\sigma\omega}\Big)\nonumber\\
\times P\Big((a_{22})_{\omega'}|(b_{11},b_{22})_{\sigma\omega}\Big).
\label{eq:Pa11[22]a22}
\ea

Equation (\ref{eq:Pa11[22]a22}) corresponds to the {\em Principle
IV} proposed in \cite{as97.2}.

\smallskip

\subsection{Avoiding to multiply causal links needlessly}

Applying "Occam's razor" RNL tries to account for the phenomena
without multiplying causal links beyond necessity, and assumes:

\ba
P\Big((a_{11[21]})_{\sigma'}|(b_{11},b_{21})_{\sigma\omega}\Big)\nonumber\\
= P\Big((a_{11[21]})_{\sigma'}|(b_{11},b_{21})_{(-\sigma)\omega}\Big)\nonumber\\
=P\Big((a_{11[21]})_{\sigma'}|(b_{21})_{\omega}\Big).
\label{eq:cpr1}
\ea

Eq. (\ref{eq:cpr1}) is an straightforward application of {\em
Principle III} in \cite{as97.2}, and can be further extended in a
natural way through the following two arrays of equalities:

\ba
P\Big((a_{11[22]})_{\sigma'}|(b_{11},b_{22})_{\sigma\omega}\Big)\nonumber\\
= P\Big((a_{11[22]})_{\sigma'}|(b_{11},b_{22})_{(-\sigma)\omega}\Big)\nonumber\\
=P\Big((a_{11[22]})_{\sigma'}|(b_{22})_{\omega}\Big).
\label{eq:cpr2}
\ea

\ba
P\Big((a_{22[11]})_{\omega'}|(b_{11},b_{22})_{\sigma\omega}\Big)\nonumber\\
= P\Big((a_{22[11]})_{\omega'}|(b_{11},b_{22})_{\sigma(-\omega})\Big)\nonumber\\
=P\Big((a_{22[11]})_{\omega'}|(b_{11},b_{21})_{\sigma\nu}\Big)\nonumber\\
= P\Big((a_{22[11]})_{\omega'}|(b_{11},b_{21})_{\sigma(-\nu)}\Big)\nonumber\\
=P\Big((a_{22[11]})_{\omega'}|(b_{11})_{\sigma}\Big).
\label{eq:cpr3}
\ea

\smallskip

\subsection{The 2 {\em non-before} impacts Theorem}

Substituting Eq. (\ref{eq:cpr2}) and (\ref{eq:cpr3}) into
(\ref{eq:Pa11[22]a22}) the proof of the 2 {\em non-before} impacts
Theorem 3.3 in \cite{as97.2} can be easily repeated to obtain:

\ba
E(a_{11[22]},a_{22})=E(b_{11},b_{22})\nonumber\\
\times E(a_{11[22]},b_{22})E(b_{11},a_{22}).
\label{eq:Ea11[22]a22}
\ea

Substitutions according Eq. (\ref{eq:Eb11b22}) and
(\ref{eq:Ea11[22]b22}) lead to:

\ba
E(a_{11[22]},a_{22})=0.
\label{eq:E'a11[22]a22}
\ea

\smallskip

\subsection{Timing $(a_{11[21]}$, $a_{22})$, e.g., Series 3.}

Time series 3 clearly corresponds to an experiment in which the
impact on BS$_{11}$ is a {\em non-before} one with relation to the
impact on BS$_{21}$, and the impact on BS$_{22}$ is a {\em
non-before} one with relation to the impact on BS$_{11}$.

Application of the rule expressed in Eq. (\ref{eq:Pa11[22]a22}) to
this case yields
\ba
P(a_{11[21]},a_{22})_{\sigma'\omega'}=\sum_{\sigma,\omega}P(b_{11},b_{21})_{\sigma\omega}\nonumber\\
\times P\Big((a_{11[21]})_{\sigma'}|(b_{11},b_{21})_{\sigma\omega}\Big)\nonumber\\
\times P\Big((a_{22})_{\omega'}|(b_{11},b_{21})_{\sigma\omega}\Big),
\label{eq:Pa11[21]a22}
\ea

and taking account of Eq. (\ref{eq:cpr1}) and (\ref{eq:cpr3}), one
gets the corresponding the 2 {\em non-before} impacts theorem:

\ba
E(a_{11[21]},a_{22})=E(b_{11},b_{21})\nonumber\\
\times E(a_{11[21]},b_{21})E(b_{11},a_{22}).
\label{eq:Ea11[21]a22}
\ea

Then substitutions according to Eq.(\ref{eq:Eb11b21}),
(\ref{eq:Ea11[21]b21}) and (\ref{eq:Ea11[22]b22}) yield:

\ba
E(a_{11[21]},a_{22})=0.
\label{eq:E'a11[21]a22}
\ea

\smallskip

\section{Other possible versions of RNL}

To this point we would like to stress that in case of the
experiment $(a_{11[21]}$, $a_{22})$ one is not led into absurdities
if one assumes a dependence of the values $(a_{22})_{\omega'}$ on
the values $(a_{11[21]})_{\sigma}$, for the values
$(a_{11[21]})_{\sigma}$ are assumed to depend on
$(b_{21})_{\omega}$, and not on $(a_{22})_{\omega'}$.\\

Therefore a multisimultaneity theory in which it holds that

\ba
P(a_{11[21]},a_{22})_{\sigma\omega}
=P^{QM}(u_{11},u_{22})_{\sigma\omega},
\label{eq:P'a11[21]a22}
\ea

cannot be excluded in principle, at least at the present stage of
analysis. Obviously, this would mean to assume (apparently without
necessity) a dependence of the value $(a_{22})_{\omega'}$ on the
value $(b_{21})_{\omega}$ through the bias of the twofold
dependence of $(a_{22})_{\omega'}$ on $(a_{11})_{\sigma'}$ and
$(a_{11})_{\sigma'}$ on $(b_{21})_{\omega}$. Accordingly Eq.
(\ref{eq:cpr3}) would fail, and neither theorem
(\ref{eq:Ea11[22]a22}) follows from relation
(\ref{eq:Pa11[22]a22}), nor theorem (\ref{eq:Ea11[21]a22}) from
relation (\ref{eq:Pa11[21]a22}).\\

Furthermore, the version of RNL presented in Section \ref{se:RNL}
assumes that the joint probabilities in experiments $(b_{11}$,
$a_{22})$, $(a_{11[22]}$, $a_{22})$ and $(a_{11[21]}$, $a_{22})$,
do not depend on whether the impact on BS$_{21}$ is a $b_{21}$ or
an $a_{21}$ one. The possibility of an alternative
multisimultaneity theory with esthetically more appealing rules has
been suggested in \cite{as97.1}.

\section{Real experiments}

A real experiment can be carried out arranging the setup used in
\cite{jrpt90} in order that one of the photons impacts on a second
beam-splitter before it is getting detected. For the values:

\ba
\phi_{11}=45^{\circ},\phi_{21}=-45^{\circ},\phi_{22}=90^{\circ},
\label{eq:re1}
\ea

Eq. (\ref{eq:Eu11u22}) and Eq. (\ref{eq:E'a11[21]a22}) yield the
predictions:

\ba
E^{QM}(u_{11},u_{22})=1\nonumber\\
E(a_{11[21]},a_{22})=0.
\label{eq:re2}
\ea

Hence, for Time Ordering 3 and settings according to (\ref{eq:re1})
the experiment represented in Fig. 1 allow us to decide between QM
and the version of RNL proposed in Section \ref{se:RNL} through
determining the experimental quantity:

\ba
E=\frac{\sum_{\sigma,\omega}\sigma\omega R_{\sigma\omega}}
{\sum_{\sigma,\omega}R_{\sigma\omega}},
\label{eq:Ere}
\ea

where $R_{\sigma\omega}$ are the four measured coincidence counts
in the detectors.\\

However the experiment does not allow us to decide between QM and
other versions of RNL based on (\ref{eq:P'a11[21]a22}).\\

\section{Conclusion}

We have discussed an experiment with successive impacts and
beam-splitters at rest which makes it possible to test Quantum
Mechanics vs Multisimultaneity theories. Although the experiment
requires only minor variations of standard setups, it has not yet
been carried out. If the results uphold QM one had taken an
important bifurcation on the Multisimultaneity road: a particular
version of RNL had been ruled out, and one should follow other
possible ones at the price of multiplying causal links; moreover
since the experiment fulfills the conditions for both first order
interferences and entanglement, it had offered a nice confirmation
of the superposition principle in a new situation. If the results
contradict Quantum Mechanics superluminal nonlocality and
relativity had unified into Multisimultaneity. In both cases the
experiment promises interesting information.

\section*{Acknowledgements}
I thank John Rarity and Paul Tapster (DRA, Malvern), Valerio
Scarani (EPF Lausanne), Juleon Schins (University of Twente) and
Harald Weinfurter (University of Innsbruck) for stimulating
discussions, and the L\'eman and Odier Foundations for financial
support.

\end{document}